\documentclass[a4paper]{article}

\usepackage{INTERSPEECH2022}
\usepackage{algorithm}
\usepackage[noend]{algpseudocode}
\usepackage{siunitx}
\usepackage{url}
\usepackage{pifont}
\usepackage{xcolor}
\usepackage{comment}
\usepackage{multirow, makecell}
\usepackage{enumitem}

\title{From Simulated Mixtures to Simulated Conversations \\as Training Data for End-to-End Neural Diarization}
\name{Federico Landini$^1$, Alicia Lozano-Diez$^2$, Mireia Diez$^1$, Luk\'a\v{s} Burget$^1$}
\address{
  $^1$Brno University of Technology, Faculty of Information Technology, Speech@FIT, Czechia\\
  $^2$AUDIAS (Audio, Data Intelligence and Speech), Universidad Aut\'onoma de Madrid, Spain}
\email{\{landini,mireia,burget\}@fit.vutbr.cz, alicia.lozano@uam.es}

\begin{document}

\maketitle
\begin{abstract}
  End-to-end neural diarization (EEND) is nowadays one of the most prominent research topics in speaker diarization. EEND presents an attractive alternative to standard cascaded diarization systems since a single system is trained at once to deal with the whole diarization problem. Several EEND variants and approaches are being proposed, however, all these models require large amounts of annotated data for training but available annotated data are scarce. Thus, EEND works have used mostly simulated mixtures for training. However, simulated mixtures do not resemble real conversations in many aspects. In this work we present an alternative method for creating synthetic conversations that resemble real ones by using statistics about distributions of pauses and overlaps estimated on genuine conversations. Furthermore, we analyze the effect of the source of the statistics, different augmentations and amounts of data. We demonstrate that our approach performs substantially better than the original one, while reducing the dependence on the fine-tuning stage. Experiments are carried out on 2-speaker telephone conversations of Callhome and DIHARD 3. Together with this publication, we release our implementations of EEND and the method for creating simulated conversations.
\end{abstract}

\vspace{-0.2cm}
\section{Introduction}
\vspace{-0.1cm}

Since the introduction of end-to-end neural diarization (EEND)~\cite{fujita2019enda} and its extension to deal with arbitrary amounts of speakers~\cite{fujita2020neural,horiguchi2020end}, it has been established as a state-of-the-art alternative to the standard cascaded diarization systems based on different submodules, i.e. voice activity detection (VAD), uniform segmentation, speaker embeddings extraction, clustering and overlapped speech detection (OSD) with handling. 
EEND formulates the speaker diarization problem as a per-speaker-per-time-frame binary classification problem where a permutation-free objective is used to minimize the speech activity error for all speakers. Therefore, EEND models generate one speech activity probability output for each speaker per time-step, which effectively solves VAD, speaker labeling and OSD at once. 

Most works following the EEND principle have focused on improvements on the architecture or modeling. Some by using self-attention layers~\cite{fujita2019endb} or conformer layers~\cite{chieh2021end} instead of the original BLSTM layers for feature encoding; others have focused on more complex diarization scenarios such as its online fashion~\cite{han2021bw,xue2021onlineA} or when more than one microphone is available~\cite{horiguchi2021multi} or by improving the model iteratively using pseudo-labels~\cite{takashima2021end}. Some have used EEND together with more standard approaches by using EEND-inspired models to find overlaps among pairs of speakers in the output of a cascaded system~\cite{horiguchi2021end} or leveraging EEND's VAD performance by using an external VAD system~\cite{horiguchi2021hitachi} or combining short duration diarization outputs to produce better whole-utterance diarization~\cite{kinoshita2021integrating,kinoshita2021advances,horiguchi2021towards,kinoshita2022tight}. However, none have yet tackled one of the main aspects of EEND: training data generation.

EEND models require large amounts of training data and datasets manually annotated for diarization do not amount for thousands of hours. When presenting the first version of EEND, Yusuke et al. proposed a strategy for constructing simulated mixtures using telephone conversations from different collections and this strategy has been used with both telephony data or read books to create simulated mixtures by mixing speakers from different recordings. However, little analysis has been presented about how the simulations were devised nor what impact have the used augmentations. Furthermore, the mixtures do not resemble real conversations, specially when more than two speakers are included. In this work we revise the approach and propose an alternative method that, based on statistics from real conversations, emulates some of the attributes of natural conversations. We also analyze the impact of using different types of augmentations on the diarization performance and show that our approach performs better than the original one while significantly reducing the dependence on the fine-tuning stage.

\vspace{-0.1cm}
\section{Training data generation strategy}
\vspace{-0.1cm}

Diarization training data consist of audio recordings and their corresponding speaker segment annotations. In this section, we describe the original approach~\cite{fujita2019enda} for creating the training data to which, respecting the original terminology, we will refer as simulated mixtures (SM) as well as our approach to which we will refer as simulated conversations (SC).

\vspace{-0.2cm}
\subsection{Simulated mixtures}
\vspace{-0.1cm}

In order to create the mixtures with their corresponding annotations, Yusuke et al. have used a VAD system run on each side of each conversation in a pool of thousands of telephone conversations. Assuming a single speaker per telephone channel, this means that speech segments and their speaker labels can be gathered from the pool to construct mixtures.

To create a mixture, as many speakers as wanted in the mixture ($N_{spk}$) are sampled from the total pool. The pool is represented by $\mathcal{U} = \{ U_s \}_{s \in \mathcal{S}}$ where $U_s$ is an utterance of speaker $s$ formed by all the segments denoted by the VAD that belong to that speaker in the utterance\footnote{Please note that Yusuke et al. have referred as ``utterances'' to the segments of speech in a real conversation.}. For each selected speaker, one of their utterances is randomly sampled and $N_u$ consecutive segments selected randomly from it. 
Pauses are introduced in between the selected segments of a speaker to simulate turns in a conversation, where the length of the pause is sampled from an exponential distribution with parameter $\beta$. The resulting audio defines a channel for that speaker in the mixture. This process is repeated for all speakers in the mixture and finally the channels are summed to obtain a single utterance. This procedure is enriched by adding background noises and altering the room impulse responses (RIRs) of each speaker channel. For the sake of space we do not include the algorithm but refer the reader to Algorithm 1 of~\cite{fujita2019enda}. Note that since the channel of each speaker is generated independently, depending on the choice of $\beta$, the output can contain a high percentage of overlapped speech (usually higher than in real conversations).

\vspace{-0.2cm}
\subsection{Simulated conversations}
\vspace{-0.1cm}

One of the main concerns with SM is that each speaker in the mixture is treated independently. Although the lengths of the pauses are randomly drawn from an exponential distribution, a sensible choice as it represents the inter-arrival times between independent events,
this does not resemble the dynamics of a real conversation where speakers do not take turns independently but collaboratively. For this reason, the proposed approach to create SC uses statistics about frequencies and lengths of pauses and overlaps from real conversations. Statistics are:

\vspace{-0.1cm}
\begin{itemize}[wide=10pt]
    \setlength\itemsep{-0.2em}
    \item Pauses between consecutive same speaker segments. The histogram of the pause lengths defines the distribution $D_{=\text{speaker}}$.
    \item Number of pauses $ds$ between consecutive different speaker segments. The histogram of the pause lengths defines the distribution $D_{\neq\text{speaker}}$.
    \item Number of overlaps $ov$ between consecutive different speaker segments. The histogram of the overlap lengths defines the distribution $D_{\text{overlap}}$.
    
    \item $p = \frac{ds}{ds + ov}$ is the probability of having a pause in between two segments of different speakers.
\end{itemize}
\vspace{-0.1cm}

The proposed approach is described in Algorithm~\ref{alg:conversation}. In this case, utterances are sampled without replacement. When creating a large set of simulated conversations, this ensures that an utterance is not used more than once\footnote{In practice, the code is prepared to run until exhausting all utterances and re-start again until a given amount of times is reached or until generating a specific amount of audio.}. Furthermore, for a given utterance all segments are used as part of a single SC. In contrast, in the original approach, only a subset of the segments from each original conversation is used at a time. 

The segments (defined by the VAD labels) of the selected utterances (one per speaker) are randomly interleaved, guaranteeing that the per-speaker order is kept while assigning random order to the speaker turns of the different speakers. Then, for each pair of consecutive segments after interleaving them, a gap is defined depending on the nature of the two segments as shown in Algorithm~\ref{alg:conversation}. Pauses or overlaps lengths are sampled using the estimated distributions. In the pseudo-code, there is a single channel initialized with $0$'s and segments are added to it in the obtained order by means of $out.addFromPosition(pos, in)$ which adds the signal $in$ onto $out$ starting from position $pos$.

In the analysis that we present in this work we consider the addition of background noises and reverberation (already present in~\cite{fujita2019enda}) as augmentations. For analyzing the effect of using or not using each of these augmentations, different training datasets are generated for each combination. The same EEND model is trained on each one independently and its performance evaluated on real data. We also explored scaling the energy levels between speakers in the conversations to resemble those in real conversations but this did not impact the performance.

\setlength{\textfloatsep}{8pt}
\begin{algorithm}
\caption{Conversation simulation}\label{alg:conversation}
\hspace*{\algorithmicindent} \textbf{Input:} $\mathcal{S}$, $\mathcal{N}$, $\mathcal{I}$, $\mathcal{R}$ \Comment{\footnotesize{Set of speakers, noises, RIRs, and SNRs}} \\
\hspace*{\algorithmicindent} \hspace*{\algorithmicindent} $\mathcal{U} = \{ U_s \}_{s \in \mathcal{S}}$ \Comment{\footnotesize{Set of utterances}} \\
\hspace*{\algorithmicindent} \hspace*{\algorithmicindent} $N_{\text{spk}}$ \Comment{\footnotesize{\#speakers per conversation}} \\
\hspace*{\algorithmicindent} \hspace*{\algorithmicindent} $p, D_{=\text{speaker}}, D_{\neq\text{speaker}}, D_{\text{overlap}}$ \Comment{Estimated distributions}
\begin{algorithmic}[1]
\State $G \gets \{\}$ \Comment{\footnotesize{Dictionary with list of segments per speaker}}
\State Sample a set of $N_{\text{spk}}$ speakers $\mathcal{S}'$ from $\mathcal{S}$ 
\ForAll{$s \in \mathcal{S}'$}
    \State Sample $u$ from $U_s$ without replacement
    \State Sample $\mathbf{i}$ from $\mathcal{I}$ \Comment{RIR}
    \State $u' \gets u * \mathbf{i}$ \Comment{Reverberate all segments in the utterance}
    \State $G[s] \gets u'$
\EndFor
\State $L \gets$ Randomly interleave $G$ into a single list
\State $\mathbf{y}.addFromPosition(0, L[1]$) \Comment{Start signal with first segment}
\State $pos \gets \text{len}(L[1])$
\For{$t = 2$ to $\text{len}(L)$}
    \If{Speaker$(L[t-1]) = $ Speaker$(L[t])$}
        \State Sample $gap$ from $D_{=\text{speaker}}$
    \ElsIf{Sample of $Bernoulli(p)$ is 1}
        \State Sample $gap$ from $D_{\neq\text{speaker}}$
    \Else
        \State Sample $-gap$ from $D_{\text{overlap}}$
    \EndIf
    \State $\mathbf{y}.addFromPosition(pos + gap, L[t])$
    \State $pos \gets pos + gap + \text{len}(L[t])$
\EndFor
\State Sample $\mathbf{n}$ from $\mathcal{N}$ \Comment{Background noise}
\State Sample $r$ from $\mathcal{R}$ \Comment{SNR}
\State Determine a mixing scale $p$ from $r$, $\mathbf{y}$, and $\mathbf{n}$
\State $\mathbf{n}' \gets$ repeat $\mathbf{n}$ until reaching the length of $\mathbf{y}$
\State $\mathbf{y} \gets \mathbf{y} + p \cdot \mathbf{n}'$
\end{algorithmic}
\end{algorithm}

For the sake of making the comparison between the two approaches as fair as possible, several aspects of the data preparation follow those of the original SM approach~\cite{fujita2019enda}:

\begin{itemize}[wide=10pt]
    \setlength\itemsep{-0.2em}
    \item The set of utterances used: comprised of Switchboard-2 (phases I, II, III)~\cite{graff1998switchboard,graff1999switchboard,graff2002switchboard}, Switchboard Cellular (parts 1 and 2)~\cite{graff2001switchboard,graff2004switchboard}, and NIST Speaker Recognition Evaluation datasets (from years 2004, 2005, 2006, 2008)~\cite{nist20112006test,nist20122006test,nist20112008train,nist20112008test,nist20112005test,nist20112006train,nist20062004,nist20112005}. All the recordings are sampled at 8\,kHz and, out of 6381 speakers, 90\% are used for creating training data.
    \item The VAD used to obtain time annotations: based on time-delay neural networks and statistical pooling\footnote{http://kaldi-asr.org/models/m4}.
    \item The set of background noises and mechanism for augmenting data: 37 noises labeled as ``background'' in the MUSAN collection~\cite{snyder2015musan} are added to the signal scaled with a signal to noise ratio (SNR) selected randomly from \{5, 10, 15, 20\}.
    \item The set of room impulse responses (RIRs) used to reverberate utterances: a RIR is sampled from the collection used in~\cite{ko2017study} and with 0.5 probability used to reverberate the utterances of each speaker.
\end{itemize}
\vspace{-0.2cm}

In order to shed some light on how the proposed SC resemble real conversations more than SM, some statistics about the recordings are presented in Table~\ref{tab:datasets_stats} where we see that SC is more similar to real sets in terms of percentages of silence, speech from a single speaker and overlap. We selected a subset of the SC to match the amount of hours using with SM in previous works. An analysis on the performance with different amounts of hours of data is presented in~\ref{sec:dataamount}. The code for generating SC can be found in https://github.com/BUTSpeechFIT/EEND\_dataprep

\begin{table}[!tb]
\setlength{\tabcolsep}{3pt} 
\centering
\caption{Statistics for synthetic and real datasets. All listed sets have only 2 speakers per recording.}
\vspace{-0.2cm}
\label{tab:datasets_stats}
\setlength{\tabcolsep}{2pt} 
  \begin{tabular}{@{}
                  lcccccc
                  @{}}
  \toprule
        Dataset       & 
        \multirow{2}{*}{\#files} & Total & Average &
        \multicolumn{3}{c}{Average \%} \\
        & & audio (h) & dur. (s) & sil. & 1spk & over. \\
  \midrule
  SM ($\beta$=2)& 100k &  2480 & 89.30 & 18.61 & 49.62 & 31.77 \\ 
  SC & 25k & 2480 & 356.15 & 12.80 & 78.83 & 8.37 \\
  \midrule
  CH Part1  & 155  & 3.19 & 74.02 & 9.05 & 77.90 & 13.05 \\
  CH Part2  & 148  & 2.97 & 72.14 & 9.84 & 78.34 & 11.82 \\
  DH3 dev & 61 & 10.17 & 599.95 & 10.56 & 77.27 & 12.17 \\
  DH3 eval & 61 & 10.17 & 599.95 & 10.89 & 78.62 & 10.49 \\
   \bottomrule
  \end{tabular}
\end{table}

\vspace{-0.1cm}
\section{Experimental setup}
\vspace{-0.1cm}

\subsection{Diarization models}
\vspace{-0.1cm}

The experiments were performed using the self-attention EEND with encoder-decoder attractors for showing superior performance in previous works. In all cases the architecture used was exactly the same as that described in~\cite{horiguchi2020end}\footnote{We refer the reader to~\cite{horiguchi2020end} for a scheme of the model.}. For the sake of making the code more efficient, we used our PyTorch implementation\footnote{https://github.com/BUTSpeechFIT/EEND}. 15 consecutive 23-dimensional log-scaled Mel-filterbanks (computed over 25\,ms every 10\,ms) are stacked to produce 345-dimensional features every 100ms. These are transformed by 4 self-attention encoder blocks (with 4 attention heads each) into a sequence of 256 dimensional embeddings. These are then shuffled in time and fed into the LSTM-based encoder-decoder module which decodes as many attractors as speakers are predicted. A binary linear classifier is used to obtain speech activity probabilities for each speaker (represented by an attractor) at each time (represented by an embedding). 

During inference time, classifiers' outputs are thresholded at 0.5 to determine speech activities. To ease the comparison, this parameter was not tuned in any experiment but tuning it could lead to better results. 
Each training was run for 100 epochs on a single GPU. The batch size was set to 64 or 32 with 100000 or 200000 minibatch updates of warm up respectively. Following~\cite{horiguchi2020end}, the Adam optimizer~\cite{kingma2014adam} was used and scheduled with noam~\cite{vaswani2017attention}. For fine-tuning on a development set, 100 epochs were run and the Adam optimizer was used with learning rate $10^{-5}$. For the inference as well as for obtaining the model from which to fine-tune, the models from the last 10 epochs were averaged. During training and fine-tuning phases, batches are formed by sequences of 500 Mel-filterbank outputs, corresponding to 50\,s. During inference, the full recordings are fed to the network one at a time. 

Diarization performance is evaluated in terms of diarization error rate (DER) as defined by NIST~\cite{NISTRT}.

\vspace{-0.2cm}
\subsection{Data}
\vspace{-0.1cm}

Two real telephone conversations datasets were used to report results. First, the 2000 NIST Speaker Recognition Evaluation~\cite{przybocki2001nist} dataset, usually referred as ``Callhome''~\cite{NISTSRE2000evalplan}. 
We report results on the subset of 2-speaker conversations using the standard Callhome partition\footnote{Each of these sets are listed in https://github.com/BUTSpeechFIT/CALLHOME\_sublists}. We will refer to the parts as CH1-2spk and CH2-2spk. Results on Callhome consider all speech (including overlap segments) for evaluation with a forgiveness collar of 0.25\,s. Second, the CTS domain from the recent Third DIHARD Challenge~\cite{ryant2020third}, which consists of previously unpublished telephone conversations from the Fisher collection. Both development and evaluation sets consist of 61 10-minute recordings each (full set). We will refer to the sets as DH-dev and DH-eval. Originally 8\,kHz signals, they were upsampled to 16\,kHz for the challenge and downsampled to 8\,kHz to be used in this work. As usual on DIHARD, all speech is evaluated with a collar of 0\,s.

\vspace{-0.1cm}
\section{Results}
\vspace{-0.1cm}

\subsection{SC augmentations analysis}
\vspace{-0.1cm}

\begin{figure}[t]
  \centering
  \hspace*{-0.5cm}
  \includegraphics[width=1\linewidth]{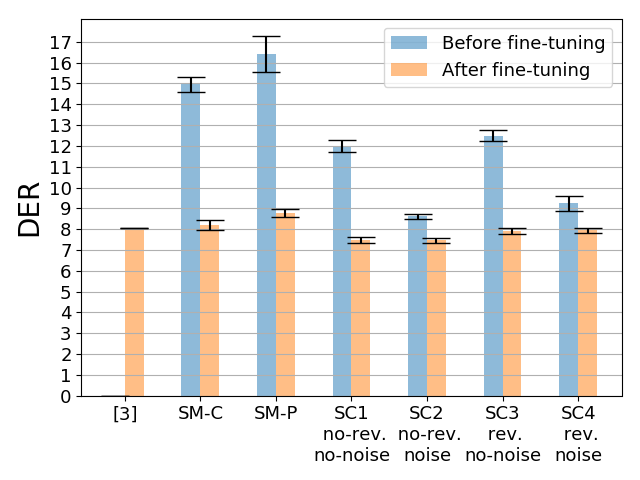}
  \vspace{-0.4cm}
  \caption{DER (\%) comparison for SM (C stands for Chainer and P for PyTorch) and SC options on CH2-2spk and fine-tuning to CH1-2spk. \cite{horiguchi2020end} is a single run. All other experiments were repeated 5 times and we show means and error bars.}
  \label{fig:v1_callhome_part2_2spk}
  \vspace*{-0.1cm}
\end{figure}

We evaluated all combinations of the augmentations, namely reverberate and add noises. For this analysis, statistics to generate SC were estimated on DH-dev. Figure~\ref{fig:v1_callhome_part2_2spk} presents results on recordings with 2 speakers from Callhome Part 2 comparing the best published result with the original Chainer implementation~\cite{horiguchi2020end}, our runs with the Chainer implementation, our PyTorch implementation with SM and the PyTorch implementation with the different SC options. Our runs with Chainer correspond with the result in~\cite{horiguchi2020end}. 
The results with PyTorch are slightly worse, note that there might be small implementation differences between Chainer and PyTorch and that the same hyper-parameters tuned for training with Chainer were used with the PyTorch implementation to ease the comparison and that adjusting them could lead to further improvement.

Using any of the SC options for generating training data outperforms using SM. Even more, the performance of the models trained using the best SC option (SC 2) is close to the results obtained with the models trained on SM after fine-tuning on the real sets. 
The models trained on the best SC option can, in turn, be further improved by means of fine-tuning, significantly outperforming those trained on SM, showing that a model trained on better quality SC data not only performs better but can also be still leveraged in combination with fine-tuning to real data.

When comparing the four options, the main gains are observed when adding background noises; as expected, this simulates noisier conditions and adds variability to the set. Reverberating the signals with the default parameters actually harms the performance. We hypothesize that the effect might be different in other scenarios such as meetings or interviews where speakers do not have specific microphones close to the mouth such as in telephone conversations. Furthermore, the choice of RIRs could be narrowed down to take into account only rooms that resemble those in real applications. These aspects need to be further studied, especially for the more diverse scenarios seen in wide-band data. Results in following sections are obtained using the best result with SC 2 and SM (P).

\vspace{-0.2cm}
\subsection{DER breakdown analysis}
\vspace{-0.1cm}

Table~\ref{tab:DER_breakdown} presents a detailed comparison of the errors from SM (P) and SC 2. In terms of VAD, both systems perform very similarly before and after fine-tuning. However, larger differences can be seen in terms of OSD. When using SC for training, the model makes considerably less false alarms for OSD, specially before fine-tuning. This can be explained by the larger percentage of overlap seen in SM (Table~\ref{tab:datasets_stats}). 
Mechanisms for favoring slightly higher percentages of overlap when creating SC are left for future studies.

\begin{table}[th]
  \caption{Error analysis before and after fine-tuning on recordings with 2 speakers of CH2-2spk.}
  \vspace{-0.2cm}
  \label{tab:DER_breakdown}
  \setlength{\tabcolsep}{3pt} 
  \centering
  \begin{tabular}{@{}
                  l
                  S[table-format=2.2] |
                  S[table-format=1.2]
                  S[table-format=1.2]
                  S[table-format=1.2] |
                  S[table-format=1.2]
                  S[table-format=1.2] |
                  S[table-format=1.2]
                  S[table-format=2.2]
                  @{}}
    \toprule
    &  & \multicolumn{3}{c|}{DER breakdown} & \multicolumn{2}{c|}{VAD} & \multicolumn{2}{c}{OSD} \\
    System & \multicolumn{1}{c|}{DER} & \multicolumn{1}{c}{Miss} & \multicolumn{1}{c}{FA} & \multicolumn{1}{c|}{Conf.} & \multicolumn{1}{c}{Miss} & \multicolumn{1}{c|}{FA} & \multicolumn{1}{c}{Miss} & \multicolumn{1}{c}{FA} \\
    \midrule
    SM (P) & 15.09 & 2.83 & 8.24 & 4.01 & 0.48 & 7.70 & 4.84 & 10.71 \\
    \hspace{0.2cm}+ FT & 8.44 & 5.23 & 2.32 & 0.90 & 3.39 & 4.06 & 6.20 & 4.22 \\
    \midrule
        SC 2 & 8.64 & 3.11 & 4.84 & 0.69 & 0.49 & 7.53 & 4.68 & 9.03 \\
    \hspace{0.2cm}+ FT & 7.28 & 4.72 & 1.98 & 0.58 & 3.23 & 4.03 & 6.02 & 3.82 \\

    \bottomrule
  \end{tabular}
  \vspace*{-0.2cm}
\end{table}

\subsection{Statistics source and fine-tuning analysis}
\vspace{-0.1cm}

To analyze the effect of the set used to estimate the statistics for creating SC data, we made a comparison using either DH-dev or CH1. As seen in Table~\ref{tab:statistics_source}, the effect of the set used for the estimation of the statistics is not large. Both datasets present several recordings that amount to few hours of speech which is enough to have a reasonable amount of data to estimate the statistics. At the same time, the fine-tuning step improves the performance on both datasets suggesting that real conversations still differ from SC, leaving room for improving SC quality.

\begin{table}[th]
  \caption{DER (\%) for models trained with SC 2 using statistics estimated on DH-dev or CH1. Fine-tuning for each test set is done in the corresponding dev set. Numbers in gray denote results where the test data were used for training. In underlined results, test data were used to compute the statistics.}
  \vspace{-0.2cm}
  \label{tab:statistics_source}
  \centering
  \begin{tabular}{@{}
                  l |
                  S[table-format=2.2] 
                  S[table-format=2.2] |
                  S[table-format=2.2]
                  S[table-format=2.2] 
                  @{}}
    \toprule
    & \multicolumn{2}{c|}{Callhome 2 speakers} & \multicolumn{2}{c}{DIHARD 3 CTS full} \\
    System & \multicolumn{1}{c}{Part 1} & \multicolumn{1}{c|}{Part 2} & \multicolumn{1}{c}{dev} & \multicolumn{1}{c}{eval} \\
    \midrule
    DH stats & \textcolor{black}{8.16} & 8.64 & \textcolor{gray}{\underline{23.47}} & 22.06 \\
    \hspace{0.2cm}+ FT & \textcolor{gray}{6.20} & 7.28 & \textcolor{gray}{16.99} & 17.00 \\
    \midrule
    CH stats & \textcolor{gray}{\underline{8.26}} & 8.73 & \textcolor{black}{22.29} & 21.53 \\
    \hspace{0.2cm}+ FT & \textcolor{gray}{6.13} & 7.28 & \textcolor{gray}{17.38} & 17.14 \\
    \bottomrule
  \end{tabular}
  \vspace*{-0.1cm}
\end{table}

\vspace{-0.4cm}
\subsection{Amount of data analysis}
\vspace{-0.1cm}
\label{sec:dataamount}

\begin{figure}[t]
  \centering
  \hspace*{-0.5cm}
  \includegraphics[width=1\linewidth]{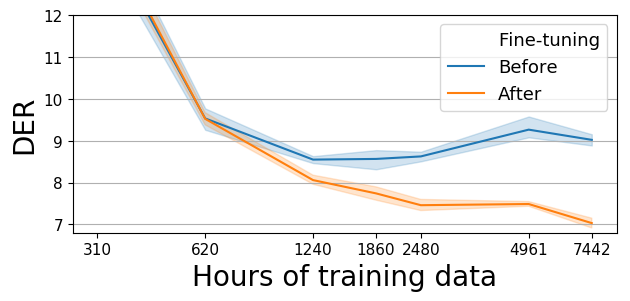}
  \vspace{-0.4cm}
  \caption{DER (\%) on CH2-2spk when training with different amounts of hours of SC 2. Each experiment is repeated 5 times to obtain the mean and confidence intervals.}
  \label{fig:confbands}
\end{figure}

One aspect to analyze is the effect that the amount of training data has on EEND performance when using SC. Figure~\ref{fig:confbands} shows that the performance when training with \textit{as little as} 310 hours degrades considerably before fine-tuning. However, using 1240 hours (half of the amount used in all other experiments) already allows for similar performance as using more data. Even more, using more data allows the model after fine-tuning to reach as little as 7.03\% DER on average and 6.8\% DER in the best run.

\vspace{-0.2cm}
\subsection{Comparison with previous results}
\vspace{-0.1cm}

In this section we compare our results to the best results published with the same architecture. Table~\ref{tab:final_results} shows that when training with SC it is possible to attain similar performance as when training with SM after fine-tuning on Callhome. This gain is smaller in the case of DIHARD CTS where fine-tuning has a larger effect. One of the aspects to consider with the architecture used is that the input is downsampled so that one output every 100\,ms is produced. When evaluating with a 0\,s forgiveness collar, such as in DIHARD, this severely impacts the results. Table~\ref{tab:final_results} presents results when evaluating at the standard downsampling of one output every 100\,ms and 50\,ms\footnote{This is done only at inference time (not during training), the system is exactly the same and cannot address the issue completely. Having finer granularity not only increases the inference time considerably but the memory requirements as well.}, showing that a considerable error reduction is possible after fine-tuning.

\begin{table}[!tb]
\centering
\caption{DER (\%) for different systems.} 
\vspace{-0.2cm}
\label{tab:final_results}
\setlength{\tabcolsep}{4pt} 
  \begin{tabular}{@{}
                  l |
                  S[table-format=2.2]
                  S[table-format=2.2] |
                  S[table-format=2.2]
                  S[table-format=2.2] |
                  S[table-format=2.2]
                  S[table-format=2.2]
                  @{}}
  \toprule
            & \multicolumn{2}{c|}{CH 2 speakers} & \multicolumn{4}{c}{DIHARD 3 CTS full} \\
    System & \multicolumn{1}{c}{Part 1} & \multicolumn{1}{c|}{Part 2} & \multicolumn{2}{c}{dev} & \multicolumn{2}{c}{eval} \\
    
  \midrule
  S.H. et al.~\cite{horiguchi2020end} & \text{--} & 8.07 & \multicolumn{2}{c}{\text{--}} & \multicolumn{2}{c}{\text{--}} \\
  \midrule
  & & & \multicolumn{4}{c}{Downsample rate (ms)} \\
  & & & \text{50} & \text{100} & \text{50} & \text{100} \\
  \midrule
  SM (P) & 13.62 & 15.09 & 25.36 & 25.46 & 22.16 & 23.58 \\
  \hspace{0.2cm}+ FT & \textcolor{gray}{7.61} & 8.44 & \textcolor{gray}{12.97} & \textcolor{gray}{17.98} & 11.99 & 17.44 \\
  \midrule
  SC 2 & 8.16 & 8.64 & \textcolor{gray}{\underline{21.16}} & \textcolor{gray}{\underline{23.47}} & 19.81 & 22.06 \\
  \hspace{0.2cm}+ FT & \textcolor{gray}{6.2} & 7.28 & \textcolor{gray}{11.68} & \textcolor{gray}{16.99} & 11.20 & 17.00 \\
  \bottomrule
  \end{tabular}
\end{table}

\vspace{-0.2cm}
\section{Conclusions}
\vspace{-0.1cm}

EEND-based diarization systems are being thoroughly explored with many different variants but they all require a vast amount of data with diarization labels for training. We presented an alternative strategy for generating training data which uses statistics of real conversations to resemble real data. Our approach significantly outperforms the original one, proving specially useful in cases where there would not be a development set for fine-tuning. However, with fine-tuning, the performance can still be leveraged showing that there is still room for improving the generation of synthetic conversations. In our future work we plan to explore the use of RIRs that resemble real scenarios and relevance of VAD and overlap related statistics. Also, we plan to further exploit our approach for creating conversations with more than 2 speakers and use it with wide-band data where there are far less hours of single-speaker recordings in conversational setups available to create synthetic conversations.

\vspace{-0.2cm}
\section{Acknowledgements}
\vspace{-0.1cm}
The work was supported by Czech Ministry of Interior project No. VJ01010108 ``ROZKAZ'', Czech National Science Foundation (GACR) project NEUREM3 No. 19-26934X and the author from the UAM was supported by project RTI2018-098091-B-I00, granted by MCIU/AEI/FEDER, UE. Computing on IT4I supercomputer was supported by the Czech Ministry of Education, Youth and Sports from the Large Infrastructures for Research, Experimental Development and Innovations project ``e-Infrastructure CZ – LM2018140''.

\bibliographystyle{IEEEtran}

\bibliography{mybib}

\end{document}